\documentclass[pre,showpacs,superscriptaddress]{revtex4}
\usepackage{graphicx}
\usepackage{dcolumn}
\usepackage{amssymb}
\usepackage{bm}
\usepackage{epsfig}
\usepackage{psfrag}

\def\deg{^{\circ}}

\def\3dots{\:\raisebox{-0.5ex}{$\stackrel{\textstyle.}{:}$}\:}
\def\beq{\begin{equation}}
\def\eeq{\end{equation}}
\def\bea{\begin{eqnarray}}
\def\eea{\end{eqnarray}}

\begin{document}

\title{Granger Causality and Cross Recurrence Plots in Rheochaos}

\author{Rajesh Ganapathy}
\affiliation{Department of Physics, Indian Institute of Science, Bangalore 560012, INDIA}
\author{Govindan Rangarajan}
\affiliation{Department of Mathematics, Indian Institute of Science, Bangalore 560012, INDIA}
\author{A.K. Sood}
\affiliation{Department of Physics, Indian Institute of Science, Bangalore 560012, INDIA}

\date{\today}
\begin{abstract}
Our stress relaxation measurements on wormlike micelles using a Rheo-SALS (rheology + small angle light scattering) apparatus allow simultaneous measurements of the stress and the scattered depolarised intensity. The latter is sensitive to orientational ordering of the micelles. To determine the presence of causal influences between the stress and the depolarised intensity time series, we have used the technique of linear and nonlinear Granger causality. We find there exists a feedback mechanism between the two time series and that the orientational order has a stronger causal effect on the stress than vice versa. We have also studied the phase space dynamics of the stress and the depolarised intensity time series using the recently developed technique of cross recurrence plots (CRPs). The presence of diagonal line structures in the CRPs unambiguously proves that the two time series share similar phase space dynamics.
\end{abstract}

\maketitle
\section{Introduction}

Wormlike micelles under shear flow are turning out to be classic examples of chaotic nonlinear dynamical systems. Wormlike micelles are long, flexible cylindrical objects formed by the self-assembly of surfactant monomers. The individual worms entangle above a critical concentration to form viscoelastic gels. Wormlike micelles relax stress through two mechanisms: reptation, analogous to polymers, and by scission and recombination \cite{catesreview}. Due to the latter process of relaxing stress, wormlike micelles show extreme shear-thinning behaviour and this shows up as a flat plateau in the flow curve (stress vs. shear rate). Cates \textit{et al.} were the first to predict this kind of flow behaviour and they attribute this to a mechanical instability of the shear banding type \cite{spenley}. In shear banding systems, the system splits into coexisting bands that support the same stress but have different average shear rates. The high shear rate band is lower in viscosity and is the nematic phase, while the low shear band has a higher viscosity and is the isotropic phase. Stress relaxation measurements in the plateau region of the flow curve for viscoelastic gels of surfactant cetyltimethyammonium tosylate (CTAT) showed chaotic dynamics \cite{ranjiniprl}. Inspite of the short data trains analysed ($\approx$ 5000 data points), owing to the low dimensional nature of the attractors, the phase space dynamics could be quantified in terms of Lyapunov exponents and fractal correlation dimensions. This phenomenon, termed rheochaos, has now been observed in many other systems \cite{ranjinieplfisher,callaghan,salmon,lootensPRL} and recently the route to rheochaos has also been observed. Ganapathy and Sood have shown that the route to rheochaos is via Type-II intermittency in stress relaxation measurements \cite{ganapathy1} for wormlike micellar gels of surfactant CTAT in the presence of salt sodium chloride (NaCl). Depolarised small angle light scattering measurements done simultaneously with flow experiments showed that the depolarised scattered intensity, sensitive to orientational order fluctuations, showed similar dynamics as that of the stress/shear rate \cite{ganapathy1}.

In the experiments reported so far \cite{callaghan,salmon}, spatial inhomogeneity always accompanies rheochaos. Spatial inhomogeneity has been captured in many recent theoretical models of rheochaos \cite{cateschaos, fieldingprl,fieldinginterface,buddhoprl}. The theoretical model relevant to the work described in this paper, proposed by Chakrabarti \textit{et al.} \cite{buddhoprl}, is a spatial extension of the homogeneous model first proposed by Hess \textit{et al.} \cite{Hess}. This model treats the system as an orientable fluid and considers the spatio-temporal evolution of the symmetric traceless nematic order parameter \cite{buddhoprl}. The equations of the motion of the stress are not fundamental but are derived from the underlying dynamics of the alignment tensor or the local orientational order parameter. An important assumption here is that the orientational order parameter equations are studied in the passive advection approximation i.e. the effect of the stress field, arising from orientational ordering, on the flow profile is ignored. Although this model ignores features specific to wormlike micelles like breaking and recombination and shear induced elongation or breaking of the micelles, it captures the essential orientable nature of the worms under flow. This model also predicts the route to rheochaos to be via spatiotemporal intermittency \cite{buddhoprl}.

Given this strong theoretical backing, it would indeed be useful to check if there exists any cause and effect relationship between the time series of stress and the depolarised scattered intensity that were measured experimentally for the CTAT + NaCl system \cite{ganapathy1}. If the passive advection approximation used in the theoretical model \cite{buddhoprl} described above is indeed justified, one should expect to find that the orientational order has a stronger causal influence on the stress than vice versa. We use the technique first developed by Granger to study linear causal effects \cite{Granger} and a recent extension of it for nonlinear time series \cite{rangarajan} to determine if there exists any causality effects in our data. In this paper we show that there exists a Granger feedback mechanism between the two time series and that the depolarised intensity or nematic ordering has a stronger influence on the stress than vice versa. Secondly, since standard nonlinear time series analysis procedures fail to quantify the dynamics of the chaotic region in intermittent data due to insufficient data points, we have used the method of cross recurrence plots (CRPs). We have analysed a data train that shows Type-II intermittency and we find that the stress and orientational order (depolarised intensity) share similar phase space behaviour in the laminar and chaotic regions.

\section{Theory}
\subsection{Linear Granger Causality}

Consider an experiment in which two time series $X(t)$ and $Y(t)$ have been simultaneously measured. If the prediction of $X(t)$ at the current time can be improved by incorporating past information from $Y(t)$, then $Y(t)$ \textit{causes} $X(t)$. This idea of causality was formalised by Granger with reference to linear regression models of stochastic processes \cite{Granger} and has numerous applications in economics \cite{Ting}. The study of causality has been also applied to medicine \cite{Tass} and nonlinear dynamical systems \cite{schiff}. Quantitatively speaking, $Y(t)$ has a causal influence on $X(t)$, if the variance of the prediction errors of $X(t)$ decreases on including information from $Y(t)$. The basis for Granger causality lies in linear prediction theory. Let $X(t)$ be a stationary process. It can be expressed using an autoregressive (AR) model where the current value of $X(t)$ depends on $m$ past values.
\beq
\label{AR_X}
X(t) = {\sum^{m}_{i=1}\alpha_{i}X(t-i)} + \epsilon_{X}(t)
\eeq
Here $\epsilon_{X}(t)$ is the error in prediction whose magnitude can be evaluated by its variance var$(\epsilon_{X}(t))$ and $\alpha_{i}$'s are the regression coefficients. If we now include information from another time series, $Y(t)$, that has also been simultaneously measured, then the prediction of the current value of $X(t)$ based on its own past and the past values of $Y(t)$ is
\beq
\label{AR_XY}
X(t) = {\sum^{m}_{i=1}a_{i}X(t-i)} + {\sum^{m}_{i=1}b_{i}Y(t-i)} + \epsilon_{X|Y}(t)
\eeq
Here $\epsilon_{X|Y}(t)$ is the prediction error in $X(t)$ given information from $Y(t)$ and $a_{i}$, $b_{i}$ are the regression coefficients. If var$(\epsilon_{X|Y}(t)) <$ var$(\epsilon_{X}(t))$, then $Y(t)$ has a causal influence on $X(t)$. Similarly, for $Y(t)$
\beq
\label{AR_Y}
Y(t) = {\sum^{m}_{i=1}\gamma_{i}Y(t-i)} + \epsilon_{Y}(t)
\eeq
\beq
\label{AR_YX}
Y(t) = {\sum^{m}_{i=1}c_{i}Y(t-i)} + {\sum^{m}_{i=1}d_{i}X(t-i)} + \epsilon_{Y|X}(t)
\eeq
If var$(\epsilon_{Y|X}(t)) <$ var$(\epsilon_{Y}(t))$, $X(t)$ has a causal influence on $Y(t)$. If var$(\epsilon_{Y|X}(t)) <$ var$(\epsilon_{Y}(t))$ and var$(\epsilon_{X|Y}(t)) <$ var$(\epsilon_{X}(t))$, then there exists a feedback mechanism between $X(t)$ and $Y(t)$.

\subsection{Extended Granger Causality}

Granger causality was developed for linear processes and a direct application on nonlinear systems might not be appropriate. Various theoretical models have been put forward to extend Granger causality for nonlinear systems. We briefly describe below the steps involved in a recent model by Chen \textit{et al.} \cite{rangarajan}.

Consider two time series $X(t)$ and $Y(t)$. Their joint dynamics can be reconstructed using the following delay vector \cite{casdagli,boccaletti}
\beq
\label{EGCI_1}
\textbf{z}(t) = (\textbf{x}(t)^{T}, \textbf{y}(t)^{T})
\eeq
where

$\textbf{x}(t) = (X(t),X(t-\tau_{1}),...,X(t-(m_{1}-1)\tau_{1}))^{T}$,

$\textbf{y}(t) = (Y(t),Y(t-\tau_{1}),...,Y(t-(m_{2}-1)\tau_{2}))^{T}$,
$m_{1}, m_{2}$ are the embedding dimensions for the two time series determined from the method of false nearest neighbors, $\tau_{1}, \tau_{2}$ are the delays determined from the first minimum of the mutual information function and the superscript $T$ stands for transpose. To investigate causal influences the flow of time is important and we set $\tau_{1} = \tau_{2} = \tau$.

In the reconstructed joint phase space of $\textbf{z}(t)$, consider a small local neighborhood of size $\epsilon$. The dynamics of this local neighborhood can be described within a linear approximation and a linear autoregressive model can be used to predict the dynamics within the neighborhood. The errors in prediction are given by $\epsilon_{Y|X}(t)$ and $\epsilon_{X|Y}(t)$. The reconstruction and the fitting procedure are now performed on the individual time series $X(t)$ and $Y(t)$ in the same local neighborhood and the errors $\epsilon_{X}(t)$ and $\epsilon_{Y}(t)$ are determined. The error ratios var$(\epsilon_{Y|X}(t))$/var$(\epsilon_{Y}(t))$ and var$(\epsilon_{X|Y}(t))$/var$(\epsilon_{X}(t))$ are computed. Since for a nonlinear system, causal inferences cannot be made by studying a single local neighborhood, the entire process described above is repeated for various regions on the attractor and the average of the error ratios are computed. The extended Granger causality index (EGCI) is defined as $\Delta_{Y\rightarrow X} = \left\langle 1 - \textup{var}(\epsilon_{X|Y}(t))/\textup{var}(\epsilon_{X}(t))\right\rangle$, where $\left\langle \cdot\right\rangle$ stands for the average. The EGCI is now computed as a function of the neighborhood size $\epsilon$. For linear systems this index will remain roughly constant as $\epsilon$ decreases, whereas for nonlinear systems, in the small $\epsilon$ limit, the EGCI reveals the true nature of the causal influence.

\subsection{Cross Recurrence Plots}

A major drawback of standard nonlinear time series analysis routines is the requirement of a long time series. One of the techniques developed to overcome this limitation is the method of Recurrence Plots (RPs) \cite{eckmann,zilbut1,zilbut2}. RPs is a tool used to visualise phase space trajectories and as the name suggests RPs represent the recurrence of trajectories in phase space, a basic feature of nonlinear dynamical systems. The first step towards generating RPs is to completely unfold the attractor. Using the Taken's embedding theorem, the phase space trajectory, $\textbf{x}(t)$, can be reconstructed by time delayed vectors of the time series $X(t)$,
\beq
\label{CRP_1}
\textbf{x}(t) = \vec{x_{i}} = (X(t),X(t-\tau_{1}),...,X(t-(m-1)\tau)).
\eeq
Here $m$ and $\tau$ are the embedding dimension as described in the previous section. The recurrence plot is defined as
\beq
\label{CRP_2}
\textbf{R}_{i,j} = \Theta(\epsilon_{i} - \left\|\vec{x_{i}} - \vec{x_{j}}\right\|)
\eeq
Here $\epsilon_{i}$ is a suitably defined cut-off distance, $\left\|\cdot\right\|$ is the Euclidean norm and $\Theta$ is the Heaviside function. If $x_{j}$ falls within distance $\epsilon_{i}$ of $x_{i}$, then the two trajectories are close by in phase space and $\textbf{R}_{i,j} = 1$, else $\textbf{R}_{i,j} = 0$. The 1's and 0's are color coded and represented as a plot.

A recent extension of RPs to bivariate time series is the method of Cross Recurrence Plots (CRPs) \cite{zilbut3,marwan1,marwan3}. Here both the time series are simultaneously delay-embedded in the same phase space and a test for closeness of the two trajectories is carried out. The CRP is defined as
\beq
\label{CRP_3}
\textbf{CR}_{i,j} = \Theta(\epsilon_{i} - \left\|\vec{x_{i}} - \vec{y_{j}}\right\|)
\eeq

RPs yield a wealth of visual information. Trajectories that share similar local dynamics show up as diagonals. Static regions in the trajectory are seen as vertical or horizontal structures. In order to be more quantitative, Zilbut and Webber have proposed the following recurrence measures \cite{zilbut1,zilbut2}.
1. \%Recurrence: This is defined as the \% of recurrence points falling within a given radius.
2. \%Determinism: This is defined as the number of recurrence points forming diagonal line structures. The more \%determinism, the more predictable is the signal. Stochastic signals have no predictability, whereas chaotic signals have short-term predictability
3. MaxLine: This is the length of the longest diagonal. This is inversely related to the maximum Lyapunov exponent which measures the rate at which phase space trajectories diverge.
4. Entropy: This is the Shannon Information entropy and is a measure of randomness of the signal. It has units of bits/bin. A periodic signal has zero entropy since all the diagonal are of the same length. A random process has a finite non-zero Shannon entropy.
5. Trend: This is a measure of the stationarity of the time series.
6. \%Laminarity: This is a measure of the \% of recurrence points forming vertical or horizontal line structures.
7. Trapping Time: This is a measure of the average length of the vertical or horizontal line structures. The last two measures were proposed by Marwan \textit{et al.} \cite{marwan2}.

\section{Experimental Methods}

The data analysed in this paper were obtained from our earlier stress relaxation experiments on the wormlike micellar system of CTAT 2wt\% + 100mM NaCl described in detail in \cite{ganapathy1}. Under controlled shear rate conditions this system shows the Type-II intermittency route to rheochaos. We briefly describe here the method that was used in our earlier work for generating the bivariate time series of stress and depolarised intensity. The experiments were performed on a MCR 300 stress controlled rheometer with SALS attachments at a temperature of 26.5$\deg$C. The experiments were carried out in a cylindrical Couette geometry with top and bottom windows made of quartz glass (inner cylinder diameter $32$mm, height $16.5$mm and gap $2$mm). A vertically polarised (V) laser beam ($\lambda$ = $658$nm and spot size $1$mm) enters the gap between the cylinders along the vorticity ($\nabla\times\textbf{v}$) direction, where $\textbf{v}$ is the velocity field. An analyser below the Couette geometry allows us to select either the vertically (referred as VV) or the horizontally polarised/depolarised scattered light (referred as VH) from the sample without disturbing the measurements. A condenser beneath the analyser collects the scattered light dominantly from a plane $6$mm above the bottom plate and forms the image on a screen in the ($\textbf{v}$, $\nabla\textbf{v}$) plane. A 8-bit colour CCD camera (Lumenera 075C, 640 x 480 pixels, maximum frame rate - 60fps) at a frame rate of 1 frame/750ms imaged the screen. About $3000$ images were grabbed for each polarisation while stress relaxation measurements were simultaneously going on. The intensity at various wavevectors from the noise filtered image was measured and a time series was generated by repeating the process over each image. The data analysed in this paper are at a fixed wave vector, $q = 0.75\mu m^{-1}$.

\section{Results}

\begin{figure}[htbp]
\includegraphics[width=0.65\textwidth]{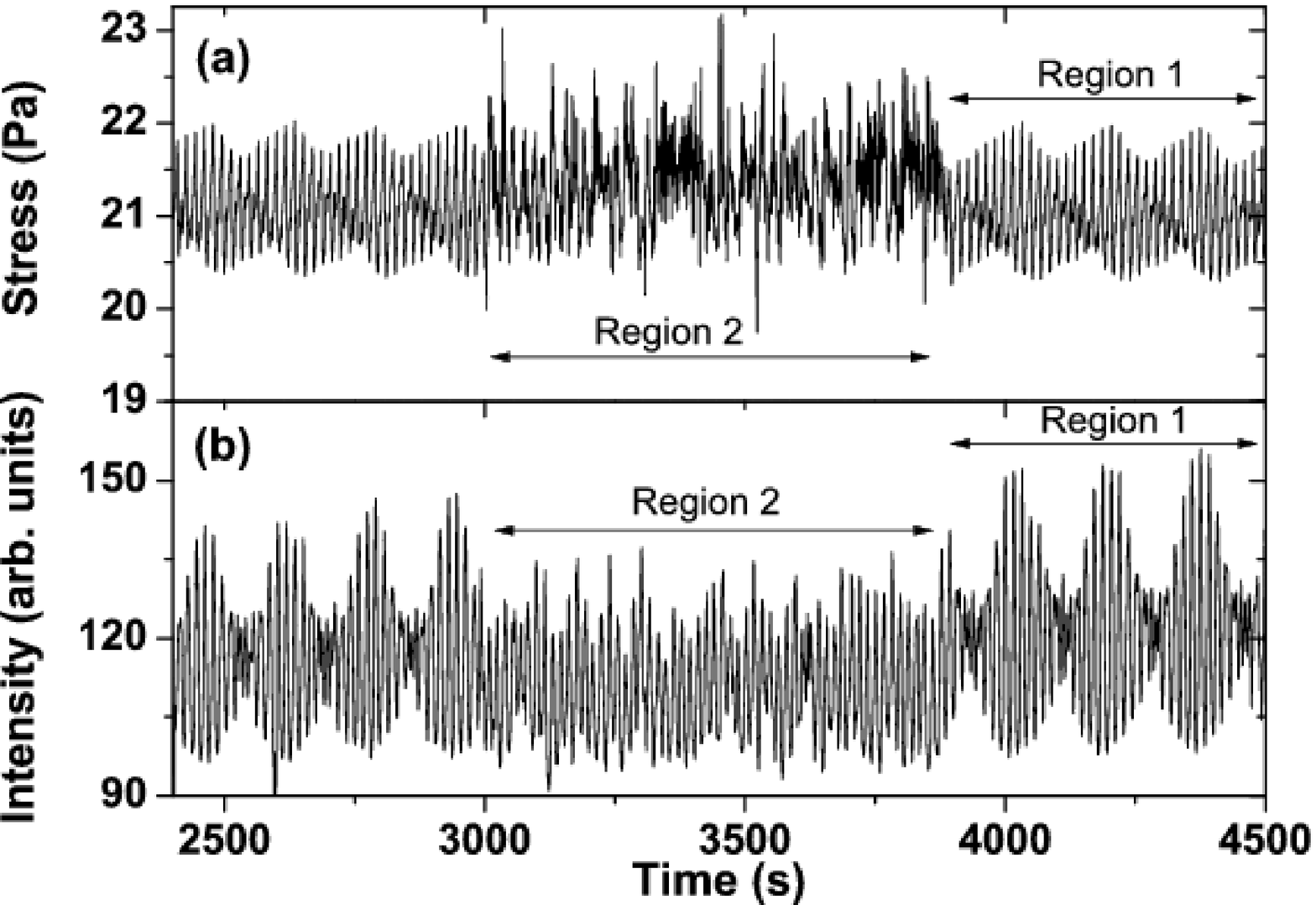}
\caption{Bivariate time series. (a) Stress time series (b) Depolarised Intensity (VH) time series obtained at a shear rate of 25$s^{-1}$.}
\label{Figure1}
\end{figure}

Figure \ref{Figure1}a \& b show the partial time series of the stress and the VH intensity at a fixed shear rate of 25 $s^{-1}$ respectively. The time series shows intermittent behaviour characterised by the presence of laminar and chaotic regions. The reason for analysing this particular data set is two-fold: 1. it would be worthwhile to check whether the two time series share similar phase space dynamics in the laminar and chaotic regions and 2. the chaotic region of the time series cannot be analysed using normal nonlinear time series analysis procedures due to the relatively short length of the chaotic regions. The laminar and chaotic regions of the time series are denoted as Region-I and Region-II and are analysed individually.

\begin{figure}[tbp]
\includegraphics[width=0.65\textwidth]{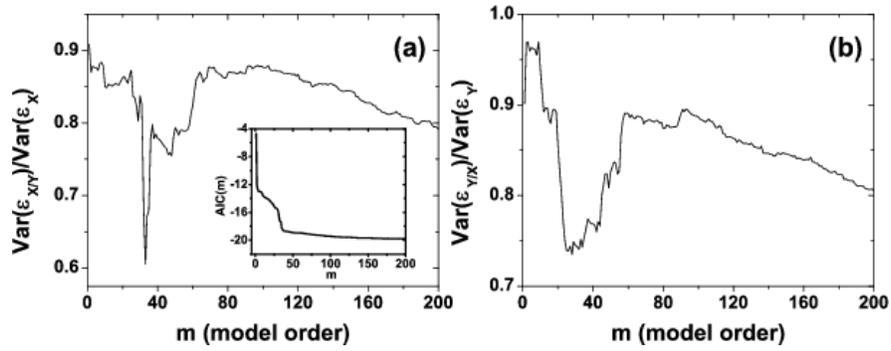}
\caption{(a) Ratio of the variance of errors for Y driving X and (b) Ratio of the variance of errors of X driving Y plotted against the regression model order $m$. Here X stand for the stress and Y for the VH intensity. The inset to (a) shows the Akaike Information Criteria (AIC) plotted against the regression model order.}
\label{Figure2}
\end{figure}

\begin{figure}[tbp]
\includegraphics[width=0.65\textwidth]{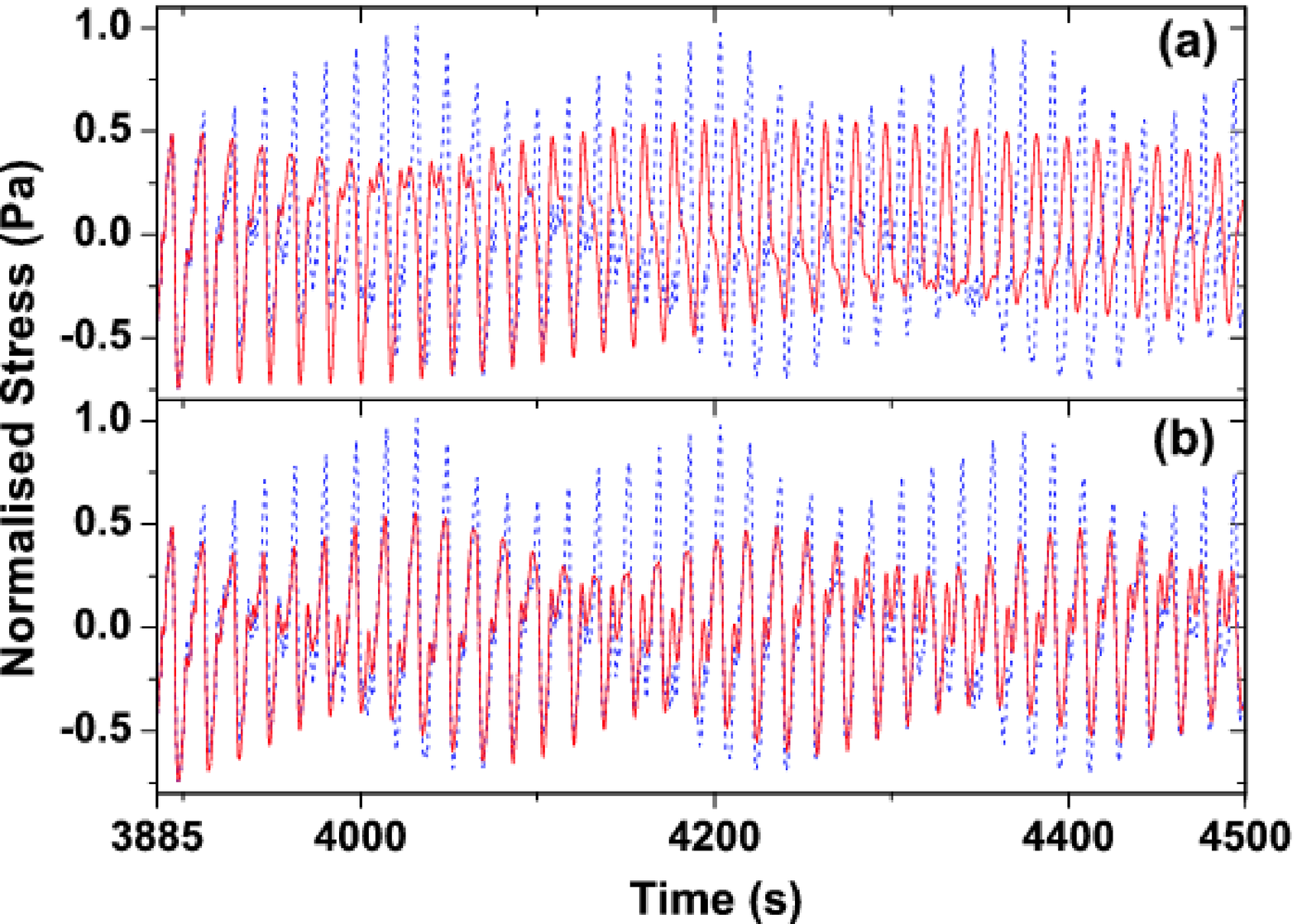}
\caption{(a) Predicted time series of stress (continuous lines) obtained without using information from VH intensity overlayed on raw time series (dashed lines). (b) Predicted time series of stress (continuous lines) obtained with using VH intensity information overlayed on raw time series (dashed lines). Regression model order $m = 33$.}
\label{Figure3}
\end{figure}

\begin{figure}[tbp]
\includegraphics[width=0.65\textwidth]{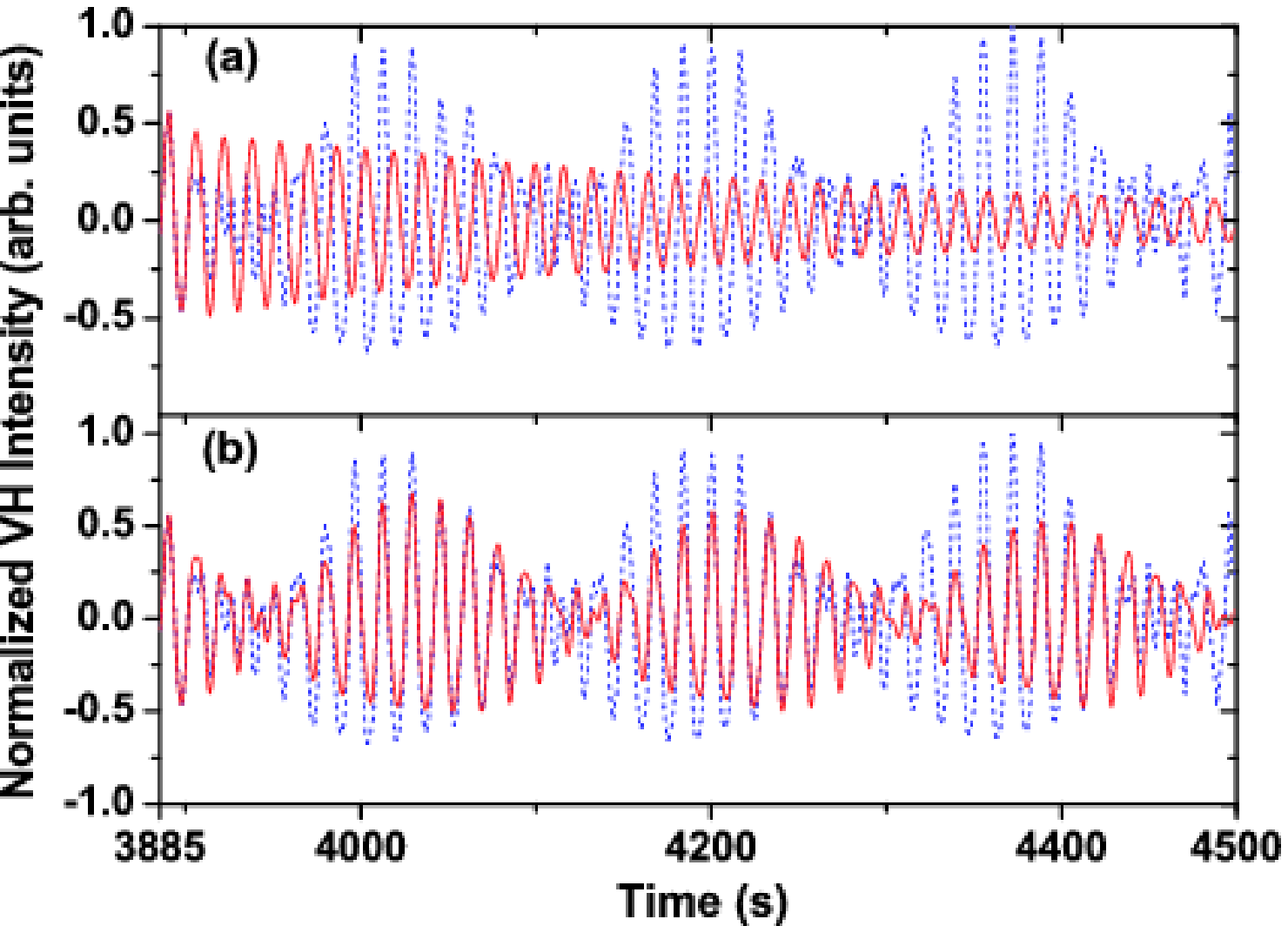}
\caption{(a) Predicted time series of VH intensity (continuous lines) obtained without using information from stress overlayed on raw time series (dashed lines). (b) Predicted time series of VH intensity (continuous lines) obtained with using stress information overlayed on raw time series (dashed lines). Regression model order $m = 33$.}
\label{Figure4}
\end{figure}

We first analyse Region-I of the time series shown in Fig. \ref{Figure1}a \& b for linear Granger causality using the procedure described in the theory section \cite{Granger}. The stress and the VH intensity time series are denoted as $X(t)$ and $Y(t)$, respectively. A linear autoregressive model fitting for each of the time series (eqns. \ref{AR_X} and \ref{AR_Y}) for increasing model order $1 < m < 200$ is done and the variance of the prediction errors of $\epsilon_{X}(t)$ and $\epsilon_{Y}(t)$ are calculated. We now incorporate information from $Y(t)$ in $X(t)$ and vice versa (eqns. \ref{AR_XY} and \ref{AR_YX}) and determine the variance of the prediction errors of $\epsilon_{X|Y}(t)$ and $\epsilon_{Y|X}(t)$ for $1 < m < 200$. The variance of the prediction errors for the mono and bivariate cases drops sharply for $1 < m < 30$ and saturates for higher values of $m$. This behaviour is typical of linear regressive models and suggests that for the experimental time series analysed here $m \approx 30$.  Fig. \ref{Figure2}a \& b show the error ratios var$(\epsilon_{X|Y}(t))$/var$(\epsilon_{X}(t))$ and var$(\epsilon_{Y|X}(t))$/var$(\epsilon_{Y}(t))$ plotted against the regression model order $m$. Over the range of $m$ analysed, the error ratios are $< 1$ implying that $X(t)$ and $Y(t)$ are better predicted when information from $Y(t)$ and $X(t)$ are incorporated in the regression equations. The error ratios are at a minimum for $m = 33$. To understand the origin of the minimum in the ratio of the variances, we have evaluated the Akaike Information Criteria (AIC) \cite{Akaike} for various model orders (inset to Fig. \ref{Figure2}a). The optimal model order estimated from the AIC yields a regression model order $m \approx 34$. This suggests that there is maximal driving between the two time series (Fig. \ref{Figure1}) for this value of $m$ and leads to a minimum in the ratio of the variances (Fig. \ref{Figure2}). For this value of $m$ the time series is predicted using the various regression coefficients that were computed from eqns. \ref{AR_X}, \ref{AR_XY}, \ref{AR_Y}, and \ref{AR_YX}. Figure \ref{Figure3} shows the predicted (continuous lines) and the raw (dashed lines) time series of the stress (Region-I of \ref{Figure1}a) overlayed on top of each other. The predicted time series compares reasonably well with the raw data when information from the VH intensity time series is used (Fig. \ref{Figure3}b) than without it (Fig. \ref{Figure3}a). Similarly, Figure \ref{Figure4} shows the predicted (continuous lines) and the raw (dashed lines) time series of the VH intensity (Region-I of \ref{Figure1}b) overlayed on top of each other. Once again, there is reasonable overlap between the predicted and raw time series when information from the stress time series is incorporated (Fig. \ref{Figure4}b). This suggests that there exists a feedback mechanism between the stress and the VH intensity in the laminar region of the time series. We have carried out the linear causality analysis for the chaotic region of the time series also. We find that the variance of the prediction errors do not saturate even for large $m$ ($m > 400$) implying that a linear regression model is inappropriate to model the chaotic regions of the time series.

\begin{figure}[tbp]
\includegraphics[width=0.6\textwidth]{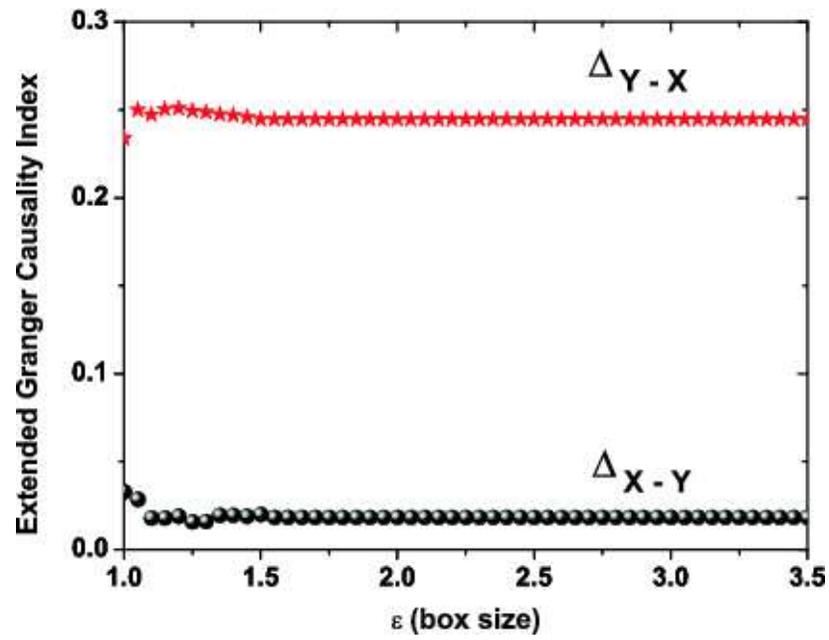}
\caption{EGCI plotted against the box size $\epsilon$. Y driving X (stars). X driving Y (circles). X stands for the stress and Y for the VH intensity.}
\label{Figure5}
\end{figure}

In order to determine the influence of nonlinear correlations in the time series on causality, we have carried out Extended Granger Causality tests \cite{rangarajan}. We describe below the results of this analysis. Following the procedure described in the theory section, we first unfold the attractor of the joint dynamics and the attractors for the individual time series in a higher dimensional phase space using the delay-embedding technique. The time delay of the two time series in the laminar region (Region-I) was found to be $\tau = 7$ and was estimated from the first minimum of the mutual information function. The embedding dimension was calculated using the method of false nearest neighbors and was found to be $m = 3$. The above two quantities were estimated using the TISEAN software \cite{TISEAN}.  For $\tau = 7$, the EGCI could not be estimated due to a lack of neighbors, however, $\tau = 1$ was found to be sufficient to obtain causal relations \cite{RuelleTakens}. The error ratios var$(\epsilon_{Y|X}(t))$/var$(\epsilon_{Y}(t))$ and var$(\epsilon_{X|Y}(t))$/var$(\epsilon_{X}(t))$ were computed for twenty different neighborhoods on the attractor and were then averaged to compute the EGCI defined earlier. The size of the neighborhood $\epsilon$ was reduced and the above procedure was repeated.
Figure \ref{Figure5} shows the EGCI plotted as a function of the box size $\epsilon$. We find $\Delta_{Y\rightarrow X} \approx 0.25$ a measure of the influence of VH intensity on the stress to be much larger than the influence of stress on the VH intensity given by $\Delta_{X\rightarrow Y} \approx 0.02$. We are also unable to go to very small $\epsilon$ due to a lack of neighbors within the neighborhood. We are unable to carry out this test in the chaotic region of the time series due to the following reason: an underlying assumption in the above procedure is that are sufficient data points within a local neighborhood to perform a linear regression analysis \cite{rangarajan}. For a high delay-embedding dimension $m = 7$ as in our case, an extremely long time series is required to carry out the Extended Granger Causality test.

\begin{table}
\begin{center}
\caption{Shows the various recurrence quantification measures for the laminar and chaotic CRPs  }
\vspace*{0.5cm}
\begin{tabular}{|c|c|c|}
\hline
Recurrence Measure &  Laminar Region & Chaotic Region \\
\hline
\%Recurrence & 2.2\% & 1.7\% \\
\%Determinism & 73.5\% & 51.2\% \\
Maxline & 10 & 21 \\
Entropy & 1.5 & 1.4 \\
Trend & 0.2 & 0.6 \\
\%Laminar & 95.1\% & 94.5\% \\
Trap Time & 3.6 & 3.9 \\
\hline
\end{tabular}
\end{center}
\end{table}

\begin{figure}[tbp]
\includegraphics[width=0.6\textwidth]{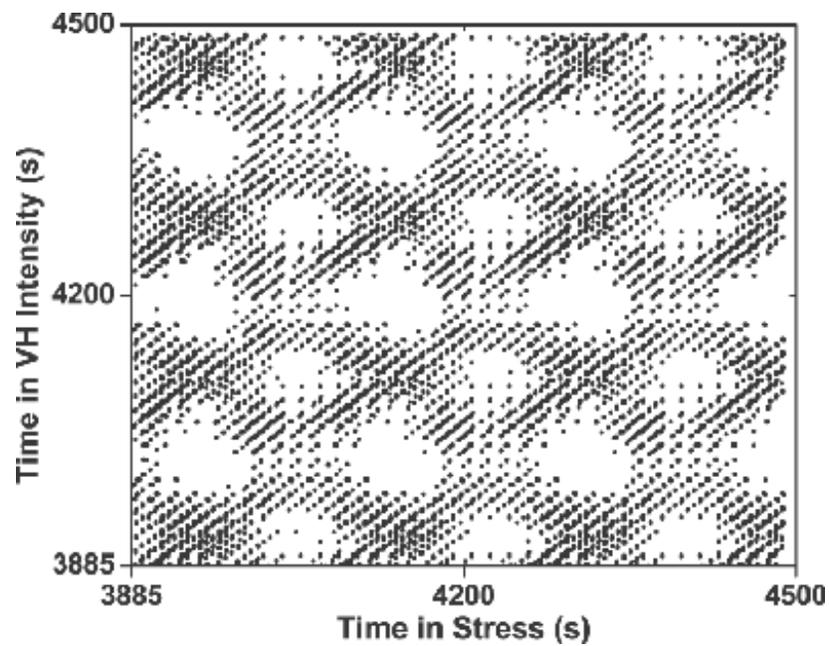}
\caption{CRP of stress and VH intensity in the laminar region.}
\label{Figure6}
\end{figure}

\begin{figure}[tbp]
\includegraphics[width=0.6\textwidth]{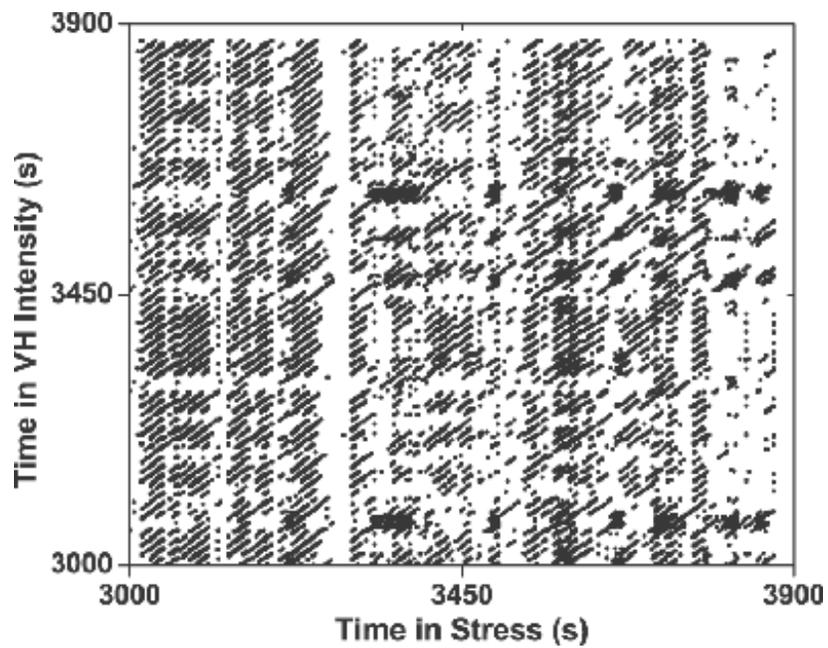}
\caption{CRP of stress and VH intensity in the chaotic region.}
\label{Figure7}
\end{figure}

We will now turn our attention CRPs. CRPs were generated and quantified using the software developed by Zilbut and Webber \cite{webberhome}. The stress and the VH intensity time series were normalized to have a zero mean and unit variance. The delay and the embedding dimension for the laminar and chaotic regions of the two time series were given in the previous section and these were used in the generation of CRPs. The radius of the neighborhood $\epsilon$ (fixed radius, Euclidean Norm) for the laminar and chaotic regions were 0.38 and 0.61 respectively. These were chosen based on the criteria that the \%Recurrence should be kept small since large $\epsilon$ might cover the entire attractor and all points will be found recurrent. Figures \ref{Figure6} \& \ref{Figure7} show the CRPs of the VH intensity and stress in the laminar and chaotic regions respectively. Inspite of the low \%Recurrence $\approx 3\%$ (Table-1) implying a small $\epsilon$, diagonal line structures can be seen. These imply that in the laminar and chaotic regions the attractors corresponding to the two time series inhabit the same regions of phase space. Not surprisingly, the \%Determinism is larger in the laminar compared to the chaotic region. The various measures defined above for quantifying CRPs are shown in Table-1.

\section{Conclusions}
To summarize, we have shown conclusively that there exists a Granger feedback mechanism between the stress and the VH intensity time series. The extended Granger causality test, apt for nonlinear dynamical systems, shows that VH intensity time series sensitive to orientational order fluctuations has a stronger influence on stress compared to the reverse. Most importantly, the extended Granger causality test puts the passive advection approximation used in the theoretical model \cite{buddhoprl} for studying the time evolution of the orientational order parameter on a firmer footing. Using CRPs we have also shown that the phase space dynamics of the bivariate time series are similar in the laminar and the chaotic regions.

\section{Acknowledgements}
GR's work was supported in part by a grant from DRDO, India and UGC SAP-DSA Phase IV. He and AKS are Honorary Faculty Members of Jawaharlal Nehru Centre for Advance Scientific Research, Bangalore, India.

\begin{thebibliography}{10}
\providecommand*{\bibinfo}[2]{#2} \providecommand*{\eprint}[1]{#1}
\providecommand*{\url}[1]{#1}

\bibitem{catesreview}
\bibinfo{author}{M. E. Cates} and \bibinfo{author}{S. J. Candau}, \bibinfo{journal}{J. Phys.: Condens. Matter} \bibinfo{volume}{2}, \bibinfo{pages}{6869} (\bibinfo{date}{1990}).

\bibitem{spenley}
\bibinfo{author}{N.~A. Spenley}, \bibinfo{author}{M. E. Cates}, and \bibinfo{author}{T. C. B. McLeish}, \bibinfo{journal}{Phys.~Rev.~Lett.} \bibinfo{volume}{\textbf{71}}, \bibinfo{pages}{939} (\bibinfo{date}{1993}).

\bibitem{ranjiniprl}
\bibinfo{author}{R.~Bandyopadhyay}, \bibinfo{author}{G. Basappa}, and \bibinfo{author}{A. K. Sood}, \bibinfo{journal}{Phys.~Rev.~Lett.} \bibinfo{volume}{\textbf{84}}, \bibinfo{pages}{2022} (\bibinfo{date}{2000}).

\bibitem{ranjinieplfisher}
\bibinfo{author}{R.~Bandyopadhyay} and \bibinfo{author}{A.~K.~Sood}, \bibinfo{journal}{Europhys.~Lett.} \bibinfo{volume}{\textbf{56}}, \bibinfo{pages}{447} (\bibinfo{date}{2001}); \bibinfo{author}{P.~Fisher}, \bibinfo{journal}{Rheol.~Acta.} \bibinfo{volume}{\textbf{39}}, \bibinfo{pages}{234} (\bibinfo{date}{2000}).

\bibitem{callaghan}
\bibinfo{author}{M.~R.~L\'opez-Gonzale\'z}, \bibinfo{author}{W. M. Holmes}, \bibinfo{author}{P. T. Callaghan}, and \bibinfo{author}{P. J. Photinos}, \bibinfo{journal}{Phys.~Rev.~Lett.} \bibinfo{volume}{\textbf{93}}, \bibinfo{pages}{268302} (\bibinfo{date}{2004}); \bibinfo{author}{W.~M.~Holmes}, \bibinfo{author}{M.~R.~L\'opez-Gonzale\'z}, and \bibinfo{author}{P.T. Callaghan}, \bibinfo{journal}{Europhys.~Lett.} \bibinfo{volume}{\textbf{64}}, \bibinfo{pages}{274} (\bibinfo{date}{2003}); \bibinfo{author}{J.~-B.~Salmon}, \bibinfo{author}{S. Manneville}, and \bibinfo{author}{A. Colin}, \bibinfo{journal}{Phys.~Rev.~E.} \bibinfo{volume}{\textbf{68}}, \bibinfo{pages}{051504} (\bibinfo{date}{2003}).

\bibitem{salmon}
\bibinfo{author}{A.~S.~Wunenburger}, \bibinfo{author}{A. Colin}, \bibinfo{author}{J. Leng}, \bibinfo{author}{A. Arneodo}, and \bibinfo{author}{D. Roux}, \bibinfo{journal}{Phys.~Rev.~Lett.} \bibinfo{volume}{\textbf{86}}, \bibinfo{pages}{1374} (\bibinfo{date}{2001}); \bibinfo{author}{J.~-B.~Salmon}, \bibinfo{author}{A. Colin}, and \bibinfo{author}{D. Roux}, \bibinfo{journal}{Phys.~Rev.~E.} \bibinfo{volume}{\textbf{66}}, \bibinfo{pages}{031505} (\bibinfo{date}{2002}); \bibinfo{author}{L.~Courbin}, \bibinfo{author}{P. Panizza}, and \bibinfo{author}{J. -B. Salmon}, \bibinfo{journal}{Phys.~Rev.~Lett.} \bibinfo{volume}{\textbf{92}}, \bibinfo{pages}{018305} (\bibinfo{date}{2004}).

\bibitem{lootensPRL}
\bibinfo{author}{D.~Lootens}, \bibinfo{author}{H. Van Damme}, and \bibinfo{author}{P. H\'ebraud}, \bibinfo{journal}{Phys.~Rev.~Lett.} \bibinfo{volume}{\textbf{90}}, \bibinfo{pages}{178301} (\bibinfo{date}{2003}).

\bibitem{ganapathy1}
\bibinfo{author}{R.~Ganapathy} and \bibinfo{author}{A.~K.~Sood}, \bibinfo{journal}{Phys.~Rev.~Lett.} \bibinfo{volume}{\textbf{96}}, \bibinfo{pages}{108301} (\bibinfo{date}{2006}).

\bibitem{cateschaos}
\bibinfo{author}{A.~Aradian} and \bibinfo{author}{M.~E.~Cates}, \bibinfo{journal}{Europhys.~Lett.} \bibinfo{volume}{\textbf{70}}, \bibinfo{pages}{397} (\bibinfo{date}{2005}).

\bibitem{fieldingprl}
\bibinfo{author}{S.~M.~Fielding} and \bibinfo{author}{P.~D.~Olmsted}, \bibinfo{journal}{Phys.~Rev.~Lett.} \bibinfo{volume}{\textbf{92}}, \bibinfo{pages}{084502} (\bibinfo{date}{2004}).

\bibitem{fieldinginterface}
\bibinfo{author}{S.~M.~Fielding} and  \bibinfo{author}{P.~D.~Olmsted}, \bibinfo{journal}{Phys.~Rev.~Lett.} \bibinfo{volume}{\textbf{96}}, \bibinfo{pages}{104502} (\bibinfo{date}{2006}).

\bibitem{buddhoprl}
\bibinfo{author}{B.~Chakrabarti}, \bibinfo{author}{M. Das}, \bibinfo{author}{C. Dasgupta}, \bibinfo{author}{S. Ramaswamy}, and \bibinfo{author}{A. K. Sood}, \bibinfo{journal}{Phys.~Rev.~Lett.} \bibinfo{volume}{\textbf{92}}, \bibinfo{pages}{055501} (\bibinfo{date}{2004}); \bibinfo{author}{M.~Das}, \bibinfo{author}{B. Chakrabarti}, \bibinfo{author}{C. Dasgupta}, \bibinfo{author}{S. Ramaswamy}, and \bibinfo{author}{A. K. Sood}, \bibinfo{journal}{Phys.~Rev.~E} \bibinfo{volume}{\textbf{71}}, \bibinfo{pages}{021707} (\bibinfo{date}{2005}).

\bibitem{Hess}
\bibinfo{author}{G. Rien\"{a}cker}, \bibinfo{author}{M. Kr\"{o}ger}, and \bibinfo{author}{S. Hess}, \bibinfo{journal}{Phys.~Rev.~E} \bibinfo{volume}{\textbf{66}}, \bibinfo{pages}{040702(R)} (\bibinfo{date}{2002}).

\bibitem{Granger}
\bibinfo{author}{C. W. J. Granger}, \bibinfo{journal}{Econometrica} \bibinfo{volume}{\textbf{37}}, \bibinfo{pages}{424} (\bibinfo{date}{1969}).

\bibitem{rangarajan}
\bibinfo{author}{Y.~Chen}, \bibinfo{author}{G. Rangarajan}, \bibinfo{author}{J. Feng}, and \bibinfo{author}{M. Ding}, \bibinfo{journal}{Phys. Lett. A} \bibinfo{volume}{\textbf{324}}, \bibinfo{pages}{26} (\bibinfo{date}{2004}).

\bibitem{Ting}
\bibinfo{author}{J. J. Ting}, \bibinfo{journal}{Physica A} \bibinfo{volume}{\textbf{324}}, \bibinfo{pages}{285} (\bibinfo{date}{2003}).

\bibitem{Tass}
\bibinfo{author}{P. Tass}, \bibinfo{author}{M. G. Rosenblum}, \bibinfo{author}{J. Weule}, \bibinfo{author}{J. Kurths}, \bibinfo{author}{A. Pikovsky}, \bibinfo{author}{J. Volkmann}, \bibinfo{author}{A. Schnitzler}, and \bibinfo{author}{H. -J. Freund}, \bibinfo{journal}{Phys. Rev. Lett.} \bibinfo{volume}{\textbf{81}}, \bibinfo{pages}{3291} (\bibinfo{date}{1998}); \bibinfo{author}{M. Le van Quyen}, \bibinfo{author}{C. Adam}, \bibinfo{author}{M. Baulac}, \bibinfo{author}{J. Martinerie}, and \bibinfo{author}{F. J. Varela}, \bibinfo{journal}{Brain Res.} \bibinfo{volume}{\textbf{792}}, \bibinfo{pages}{24} (\bibinfo{date}{1998}); \bibinfo{author}{E. Rodriguez}, \bibinfo{author}{N. George}, \bibinfo{author}{J. -P. Lachaux}, \bibinfo{author}{J. Martinerie}, \bibinfo{author}{B. Renault}, and \bibinfo{author}{F. J. Varela}, \bibinfo{journal}{Nature (London)} \bibinfo{volume}{\textbf{397}}, \bibinfo{pages}{430} (\bibinfo{date}{1999}).

\bibitem{schiff}
\bibinfo{author}{S. J. Schiff}, \bibinfo{author}{P. So}, \bibinfo{author}{T. Chang}, \bibinfo{author}{R. E. Burke}, and \bibinfo{author}{T. Sauer}, \bibinfo{journal}{Phys. Rev. E} \bibinfo{volume}{\textbf{54}}, \bibinfo{pages}{6708} (\bibinfo{date}{1996}); \bibinfo{author}{J. Arnhold}, \bibinfo{author}{P. Grassberger}, \bibinfo{author}{K. Lehnertz}, and \bibinfo{author}{C. E. Elger}, \bibinfo{journal}{Physica D} \bibinfo{volume}{\textbf{134}}, \bibinfo{pages}{419} (\bibinfo{date}{1999}); \bibinfo{author}{R. Quian Quiroga}, \bibinfo{author}{J. Arnhold}, and \bibinfo{author}{P. Grassberger}, \bibinfo{journal}{Phys. Rev. E} \bibinfo{volume}{\textbf{61}}, \bibinfo{pages}{5142} (\bibinfo{date}{2000}).

\bibitem{casdagli}
M. Casdagli, in: M. Casdagli, S. Eubank (Eds.), A dynamical systems approach to input-output systems, in Nonlinear Modeling and Forecasting, Proceedings, vol. XII, Addison-Wesley, 1992, pg. 265.

\bibitem{boccaletti}
\bibinfo{author}{S. Boccaletti}, \bibinfo{author}{D. L. Valladares}, \bibinfo{author}{L. M. Pecora}, \bibinfo{author}{H. P. Geffert}, and \bibinfo{author}{T. Carroll}, \bibinfo{journal}{Phys. Rev. E} \bibinfo{volume}{\textbf{65}}, \bibinfo{pages}{035204(R)} (\bibinfo{date}{2002}).

\bibitem{eckmann}
\bibinfo{author}{J. -P. Eckmann}, \bibinfo{author}{S. O. Kamphorst}, and \bibinfo{author}{D. Ruelle}, \bibinfo{journal}{Europhys. Lett.} \bibinfo{volume}{\textbf{5}}, \bibinfo{pages}{973} (\bibinfo{date}{1987}).

\bibitem{zilbut1}
\bibinfo{author}{C. L. Webber Jr} and \bibinfo{author}{J. P. Zilbut}, \bibinfo{journal}{J. Appl. Physiol.} \bibinfo{volume}{\textbf{76}}, \bibinfo{pages}{965} (\bibinfo{date}{1994}).

\bibitem{zilbut2}
\bibinfo{author}{J. P. Zilbut} and \bibinfo{author}{C. L. Webber Jr}, \bibinfo{journal}{Phys. Lett. A} \bibinfo{volume}{\textbf{171}}, \bibinfo{pages}{199} (\bibinfo{date}{1992}).

\bibitem{zilbut3}
\bibinfo{author}{J. P. Zilbut}, \bibinfo{author}{A. Giulani}, and \bibinfo{author}{C. L. Webber Jr.}, \bibinfo{journal}{Phys. Lett. A} \bibinfo{volume}{\textbf{246}}, \bibinfo{pages}{122} (\bibinfo{date}{1998}).

\bibitem{marwan1}
\bibinfo{author}{N. Marwan} and \bibinfo{author}{J. Kurths}, \bibinfo{journal}{Phys. Lett. A} \bibinfo{volume}{\textbf{302}}, \bibinfo{pages}{299} (\bibinfo{date}{2002}).

\bibitem{marwan3}
\bibinfo{author}{N. Marwan}, \bibinfo{author}{M. W. Trauth}, \bibinfo{author}{M. Vuille}, and \bibinfo{author}{J. Kurths}, \bibinfo{journal}{Climate Dynamics} \bibinfo{volume}{\textbf{21}}, \bibinfo{pages}{317} (\bibinfo{date}{2003}).

\bibitem{marwan2}
\bibinfo{author}{N. Marwan}, \bibinfo{author}{N. Wessel}, \bibinfo{author}{U. Meyerfeldt}, \bibinfo{author}{A. Schirdewan}, and \bibinfo{author}{J. Kurths}, \bibinfo{journal}{Phys. Rev. E} \bibinfo{volume}{\textbf{66}}, \bibinfo{pages}{026702} (\bibinfo{date}{2002}).

\bibitem{Akaike}
\bibinfo{author}{H. Akaike}, \bibinfo{journal}{IEEE Transactions on Automatic Control} \bibinfo{volume}{\textbf{19}}, \bibinfo{pages}{716} (\bibinfo{date}{1974}).

\bibitem{TISEAN}
The TISEAN pacakage is written by R. Hegger, H. Kantz and T. Schreiber and may be publicly downloaded from http://www.mpipks-dresden.mpg.de/~tisean.

\bibitem{RuelleTakens}
\bibinfo{author}{D. Ruelle} and \bibinfo{author}{F. Takens}, \bibinfo{journal}{Comm. Math. Phys.} \bibinfo{volume}{\textbf{20}}, \bibinfo{pages}{167} (\bibinfo{date}{1971}).

\bibitem{webberhome}
http://homepages.luc.edu/~cwebber/ (Chuck Webber)
\end{thebibliography}
\end{document}